\documentclass[conference]{IEEEtran}
\IEEEoverridecommandlockouts

\usepackage{cite}
\usepackage{amsmath,amssymb,amsfonts,tabularray}
\usepackage{algorithmic}
\usepackage{graphicx}
\usepackage{textcomp}
\usepackage{xcolor}
\usepackage{amssymb}
\usepackage{booktabs} 
\usepackage[a4paper, total={184mm,239mm}]{geometry}
\usepackage{algorithm}
\usepackage{algorithmic}
\usepackage{tabularx}
\usepackage{tikz}
\usepackage[colorlinks=true,linkcolor=black,urlcolor=magenta,citecolor=black]{hyperref}

\newcommand*\circled[1]{\tikz[baseline=(char.base)]{
            \node[shape=circle,fill,inner sep=0pt] (char) {\textcolor{white}{#1}};}}

\def\BibTeX{{\rm B\kern-.05em{\sc i\kern-.025em b}\kern-.08em
    T\kern-.1667em\lower.7ex\hbox{E}\kern-.125emX}}
\begin{document}

\title{NDPage: Efficient Address Translation for Near-Data Processing Architectures via Tailored Page Table}

\newcommand{\todo}[1]{\textcolor{magenta}{#1}}
\newcommand{\jiang}[1]{\textcolor{purple}{#1}}







\author{\IEEEauthorblockN{Qingcai Jiang$^*$, Buxin Tu$^*$ and Hong An}
\IEEEauthorblockA{School of Computer Science and Technology, University of Science and Technology of China, Hefei, China \\
}

\IEEEauthorblockA{Email: \{jqc, tubuxin\}@mail.ustc.edu.cn, han@ustc.edu.cn}
}

\maketitle
\def\thefootnote{*}\footnotetext{These authors contributed equally to this work.}
\begin{abstract}
Near-Data Processing (NDP) has been a promising architectural paradigm to address the memory wall problem for data-intensive applications.
Practical implementation of NDP architectures calls for system support for better programmability, where having virtual memory (VM) is critical.
Modern computing systems incorporate a 4-level page table design to support address translation in VM.
However, simply adopting an existing 4-level page table in NDP systems causes significant address translation overhead because 
(1) NDP applications generate a lot of address translations, and 
(2) the limited L1 cache in NDP systems cannot cover the accesses to page table entries (PTEs).
We extensively analyze the 4-level page table design in the NDP scenario and observe that
(1) the memory access to page table entries is highly irregular, thus cannot benefit from the L1 cache, and
(2) the last two levels of page tables are nearly fully occupied.

Based on our observations, we propose NDPage, an efficient page table design tailored for NDP systems.
The key mechanisms of NDPage are 
(1) an L1 cache bypass mechanism for PTEs that not only accelerates the memory accesses of PTEs but also prevents the pollution of PTEs in the cache system, and
(2) a flattened page table design that merges the last two levels of page tables, allowing the page table to enjoy the flexibility of a 4KB page while reducing the number of PTE accesses.

We evaluate NDPage using a variety of data-intensive workloads.
Our evaluation shows that in a single-core NDP system, NDPage improves the end-to-end performance over the state-of-the-art address translation mechanism of 14.3\%; in 4-core and 8-core NDP systems, NDPage enhances the performance of 9.8\% and 30.5\%, respectively.

\end{abstract}


\section{Introduction}
Data-intensive applications spend significant time and energy moving data between the CPU and memory, which causes the "memory wall" problem in modern computing systems~\cite{boroumand2018google}.
Recent advances in 3D-stacked memory technologies, e.g., High Bandwidth Memory (HBM)~\cite{hbm}, enable the practical implementation of \textbf{near-data processing (NDP)}, which is a promising solution to alleviate the memory wall problem.
By placing computing units within the logic layer of 3D-stacked memory, NDP systems enhance performance and energy efficiency with low-latency and energy-efficient memory accesses~\cite{mutlu2022modern}.

Despite the significant performance and energy-efficiency potential of NDP, system challenges remain a major obstacle to the practical adoption of NDP architectures.
Among these challenges, efficient support for \textbf{virtual memory} is often considered one of the most significant ones in NDP architectures~\cite{hyun2024pathfinding}.
Virtual memory allows the operating system to automatically map NDP's virtual memory addresses to the corresponding physical memory addresses.
This flexible mapping provides several benefits for the NDP system, including programmer-transparent memory management and enhanced programmability.

Supporting virtual memory comes at the cost of \textbf{address translation}, which is a performance bottleneck for data-intensive applications in modern computing systems with a 4-level page table~\cite{kanellopoulos2023victima}.
In this work, we find that the address translation overhead is even more profound in NDP systems if we simply apply the 4-level page table design in NDP systems due to two reasons:
(1) The data-intensive NDP applications introduce large amounts of irregular memory accesses, which cause frequent translation lookaside buffer (TLB) misses, thus leading to many page table walks (PTWs).
(2) The area and power constraints in NDP limit the cache size and levels in NDP systems, making it hard to effectively cache the recently used page table entries (PTEs).

\textbf{Our goal} in this work is to design an efficient address translation mechanism for NDP systems by adapting conventional page table design to the characteristics of NDP workloads.
We aim to develop a practical technique that:
(1) is highly efficient in both single-core and multi-core NDP systems,
(2) remains transparent to NDP applications, and
(3) requires only modest or no hardware modifications and ISA changes.

To this end, we present NDPage, a new, highly efficient, software-transparent address translation mechanism tailored for NDP systems.
Our two \textbf{key observations} behind NDPage are:
(1) The memory access pattern of PTEs, referred to as \textbf{metadata} throughout this paper, in NDP systems is highly irregular. This irregularity prevents metadata accesses from benefiting from the cache and significantly pollutes the cache, which negatively impacts the performance of normal data accesses.
(2) The flexibility of the 4-level radix page table is unnecessary in NDP systems, as the last two levels of page tables are often fully occupied.

Based on our observations, we devise two \textbf{key mechanisms} in NDPage:
(1) An L1 cache bypass mechanism for metadata that makes PTEs not cacheable in NDP's L1 cache system, accelerating the memory access of PTEs and preventing them from polluting normal data in the cache system.
(2) A flattened L2/L1 page table design that merges the last two levels of the page table in NDP systems to reduce the sequential page table accesses during address translation. 

We evaluate NDPage with Victima, an extended version of the Sniper simulator~\cite{carlson2011sniper,carlson2014evaluation} used by prior works~\cite{kanellopoulos2023victima, kanellopoulos2023utopia}, using 11 data-intensive applications from five diverse benchmark suites.
Our evaluation yields 2 \textbf{key results} that demonstrate NDPage’s effectiveness.
\textbf{First}, in single-core environments, NDPage improves performance by 34.4\% on average over the baseline system using a four-level radix-tree-based page table, yielding 14.3\% performance improvements compared with the second-best address translation mechanism.
\textbf{Second}, in multi-core environments, NDPage improves performance by 42.6\% (40.7\%) on average over the 4-core (8-core) baseline system using a four-level radix-tree-based page table, yielding 9.8\% (30.5\%) performance improvements compared with the second-best address translation mechanism.

This paper makes the following key contributions:
\begin{itemize}
    \item We investigate the problems of applying the conventional 4-level page table design for address translation in Near-Data Processing architectures.
    \item We propose NDPage, an address translation mechanism specifically designed for NDP architectures, based on our two observations in address translation within NDP systems.
    \item We evaluate NDPage using a diverse set of data-intensive applications and demonstrate its effectiveness in both single-core and multi-core environments.
\end{itemize}

\section{Background}

\subsection{Near-Data Processing Architecture}
\begin{figure}[htbp]
    \centering
    \centerline{\includegraphics[width=\linewidth]{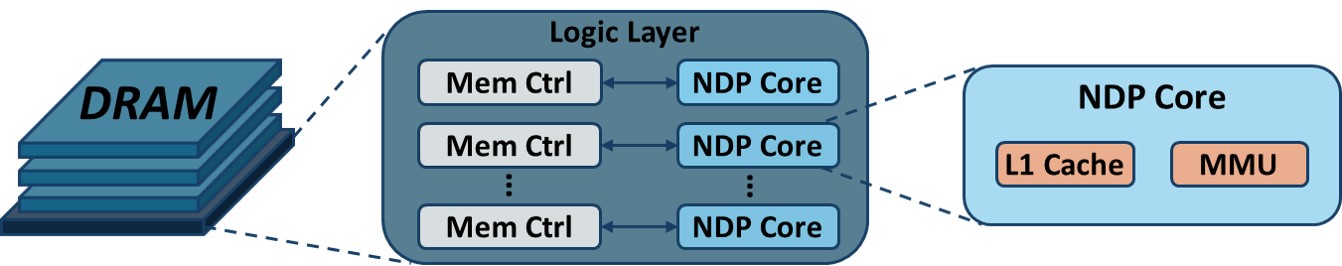}}
    \caption{Overview of NDP architecture.}
    \label{fig:NDP-architecture}
\end{figure}
Near-data processing is a promising technique to alleviate the memory wall in data-intensive applications~\cite{mutlu2022modern,giannoula2021syncron}.
By placing the computing unit (referred to as an NDP core) near the main memory, NDP cores can access data with low latency and high throughput.
Figure~\ref{fig:NDP-architecture} shows a typical NDP architecture based on 3D-stacked memories like the Hybrid Memory Cube (HMC)~\cite{hmc} and HBM~\cite{hbm,hbm2}.
In this architecture, NDP cores are placed in the logic layer of the 3D-stacked memory.
An NDP core typically has a \textbf{shallow L1 cache} due to two reasons:
(1) a strictly limited power and area budget in the logic layer,
and (2) the memory access in NDP applications usually exhibits low data locality, which cannot benefit from deep cache hierarchies~\cite{oliveira2021damov,boroumand2019conda}.
Apart from that, to support virtual memory, an NDP core incorporates a memory management unit (MMU), which will be discussed in Section~\ref{sec:vm}.

\subsection{Virtual Memory}\label{sec:vm}
Virtual memory is a fundamental concept in computer systems, designed to provide applications with the illusion of having unlimited memory~\cite{denning1970virtual, bhattacharjee2017architectural}.
This abstraction is crucial for enabling several key functionalities in computing systems, such as process isolation, data sharing, and memory protection.
To support these functionalities, \textbf{address translation} is essential for mapping virtual addresses to corresponding physical addresses.
The page table stores the metadata for these mappings.
\begin{figure}[htbp]
    \centering
    \centerline{\includegraphics[width=\linewidth]{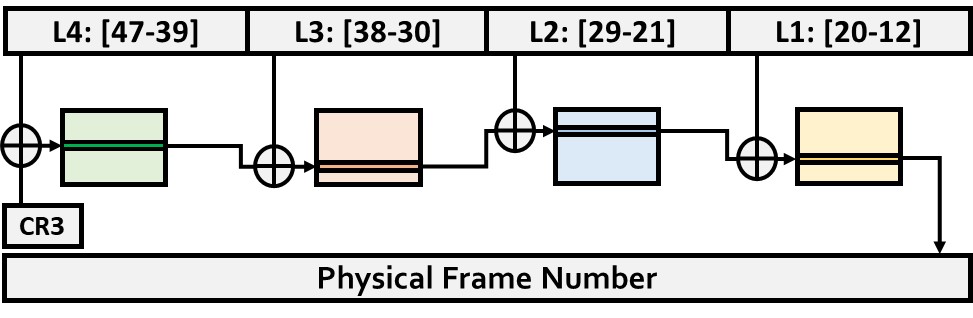}}
    \caption{Address translation in radix page table.}
    \label{fig:Radix-page-table}
\end{figure}
A commonly used page table mechanism is the radix page table~\cite{el1985intel}.
In x86-64 systems, when a virtual address needs to be mapped to a physical address, the operating system allocates memory for the page tables at each level only if they do not already exist. 
This involves allocating a 4KB page frame for each new table required. 
The flexibility provided by radix page tables results in significant space savings, which is one of their major advantages. 
Figure~\ref{fig:Radix-page-table} depicts the address translation based on a 4-level radix page table in x86-64 systems. 
The OS divides the higher 36 bits of the virtual address into four parts and executes four sequential memory accesses to page tables from level 4 to level (PL4 - PL1) to obtain the physical frame number (PFN). 

\begin{figure}[htbp]
    \centering
    \centerline{\includegraphics[width=\linewidth]{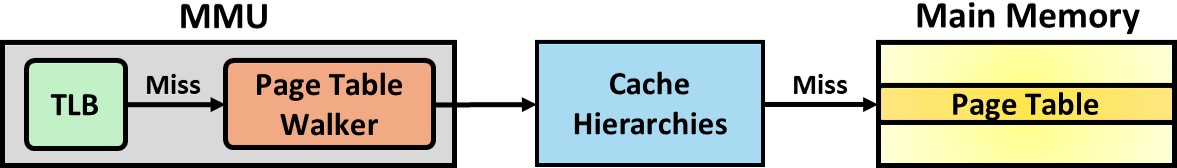}}
    \caption{Address translation workflow in modern processors.}
    \label{fig:translation-workflow}
\end{figure}
To avoid frequent access to the main memory for the page tables, virtual memory systems usually adopt the MMU hardware to accelerate address translation.
As shown in Figure~\ref{fig:translation-workflow}, an MMU has two main components:
(1) TLB, which caches recently used address translations.
(2) PTW, which initiates a page table walk upon a TLB miss, searches for PTEs in the memory hierarchies including the cache, and finally obtains the corresponding physical address. 

\section{Motivation}

Virtual memory provides an essential attraction for programmers to manage physical memory efficiently.
In contrast to maintaining a separate physical address space~\cite{gomez2021benchmarking}, virtual memory allows NDP systems to be compatible with the mainstream programming models and benefit from existing advanced memory management mechanisms~\cite{fatima2023vpim,hyun2024pathfinding}.

A fundamental aspect of virtual memory functionality is address translation, which facilitates the efficient mapping of virtual addresses to physical addresses.
To better understand the performance overhead of address translation in NDP systems, we evaluate the address translation performance metrics and compare them to those of conventional CPU systems.

\begin{figure}[htbp]
    \centering
    \centerline{\includegraphics[width=\linewidth]{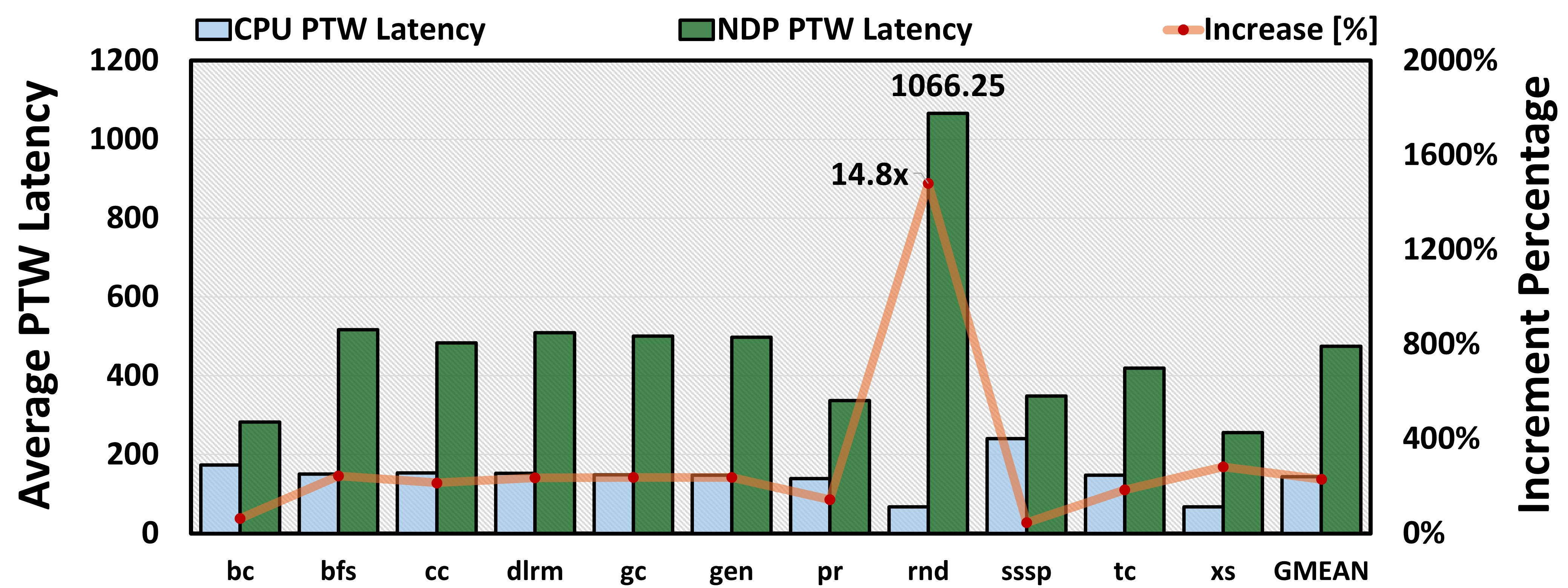}}
    \caption{Average PTW latency in 4-core systems and NDP’s PTW latency increment compared with CPU.}
    \label{fig:PTW4core}
\end{figure}

Here we take a 4-core system for example\footnote{Section~\ref{sec:methodology} shows the methodologies}. 
Figure~\ref{fig:PTW4core} demonstrates the average PTW latency in 4-core NDP and CPU systems. 
We observe that the average PTW latency in the NDP system is 474.56 cycles (up to 1066.25 cycles) across all the data-intensive applications, which is 229\% higher than in the CPU system.

\begin{figure}[htbp]
    \centering
    \centerline{\includegraphics[width=\linewidth]{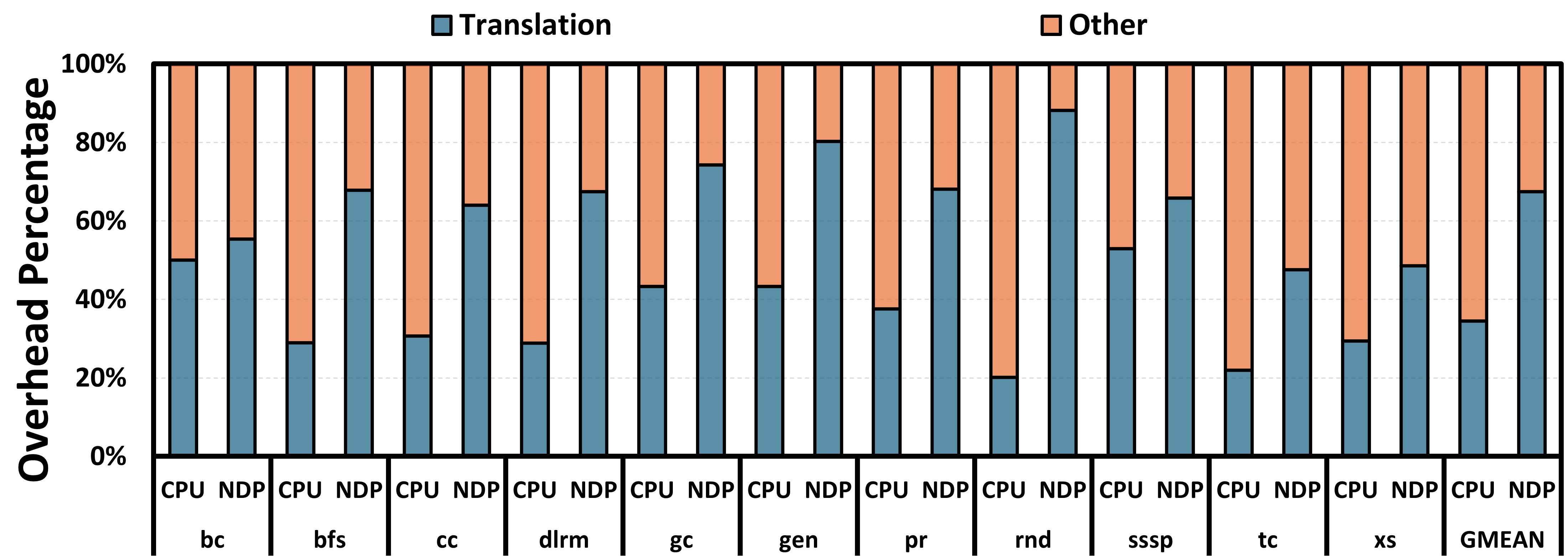}}
    \caption{Percentage of address translation overhead in 4-core systems. Blue bars denote the proportion of the address translation overhead and orange bars denote the execution time apart from address translation.}
    \label{fig:trans4core}
\end{figure}

\begin{figure}[htbp]
    \centering
    \centerline{\includegraphics[width=\linewidth]{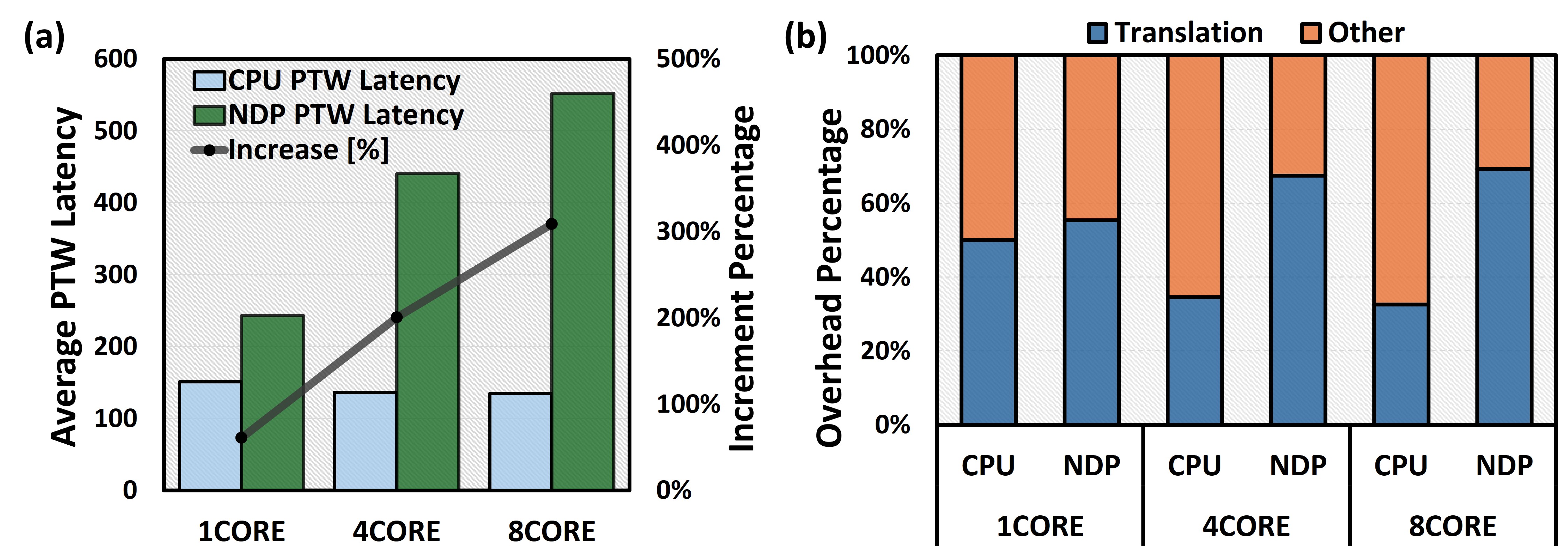}}
    \caption{As the number of cores scales, (a) shows the trend in average PTW latency, and (b) shows the trend in the average percentage of address translation overhead.}
    \label{fig:differcores}
\end{figure}

Figure~\ref{fig:trans4core} shows the percentage of total address translation overhead in 4-core NDP and CPU systems. 
We observe that the average address translation overhead in the NDP system occupies 67.1\% of the total execution time, higher than the 34.51\% observed in the CPU system. 
Furthermore, as the number of cores scales, the address translation overhead in NDP systems increases significantly. 
As Figure~\ref{fig:differcores} (a) shows, from 1 core to 8 cores, the average PTW latency in the NDP system increases from 242.85 to 551.83 cycles.
However, in the CPU system, the average PTW remains similar.
Also, as Figure~\ref{fig:differcores} (b) shows, from 1 core to 8 cores, the percentage of address translation overhead in the NDP system continues to increase, whereas in the CPU system, it also remains similar. 
We conclude that naively applying the address translation mechanism based on a 4-level radix page table on NDP systems creates huge overhead and nullifies the performance benefits of NDP architecture.

\section{Key Observations}
In this section, we study the root cause of the high address translation overhead in NDP systems.
we make 2 key observations by analyzing the memory access patterns of the PTEs and the page table occupancy:
(1) Although the memory access pattern of the normal data, i.e., the actual data that the program accesses, is irregular, the memory access pattern of \textbf{metadata}, i.e., the PTEs required to locate the \textbf{normal data}, is \textbf{even more irregular}
, and 
(2) when adopting the 4-level radix page table mechanism in NDP systems, the L2 and L1 page tables are often \textbf{fully occupied}.

\subsection{The Irregular PTE Accesses}
\label{sec:irregular-pte}
\begin{figure}[htbp]
    \centering
    \centerline{\includegraphics[width=\linewidth]{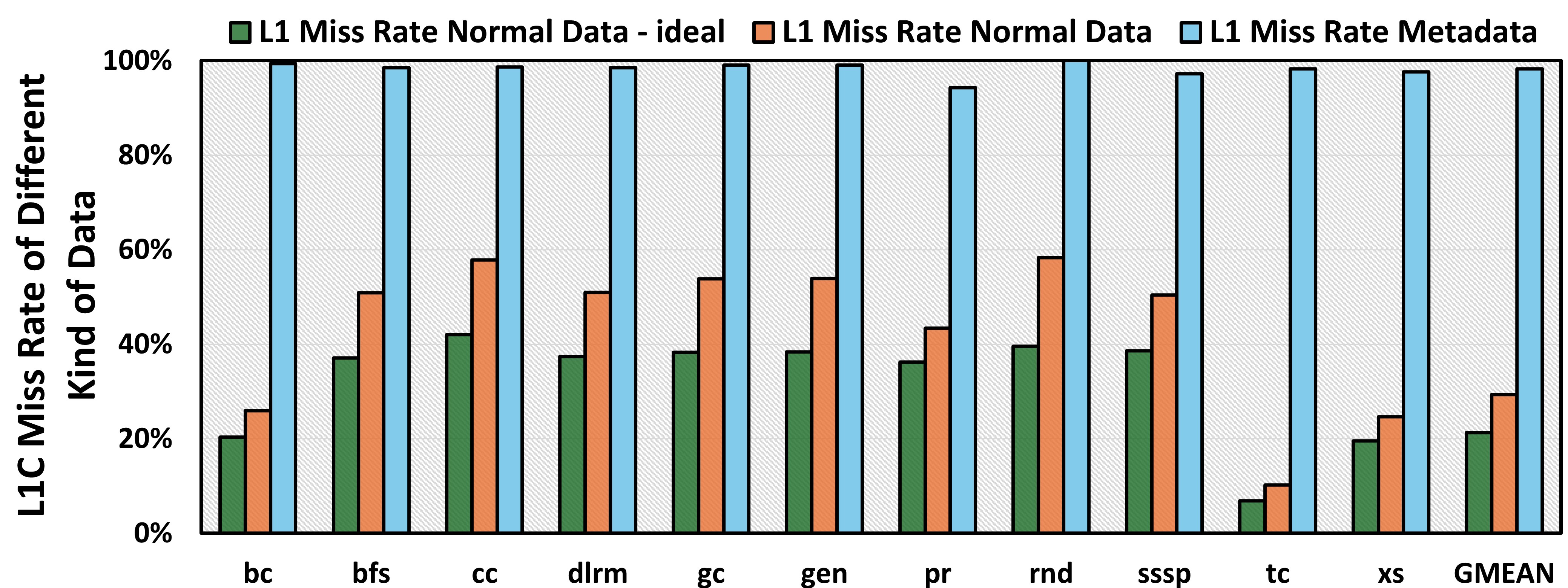}}
    \caption{Comparison of L1 data cache miss rates for normal data (ideal vs. actual) and metadata across different benchmarks in 4-core NDP systems.}
    \label{fig:missrate-cache}
\end{figure}
\begin{figure}[htbp]
    \centering
    \centerline{\includegraphics[width=\linewidth]{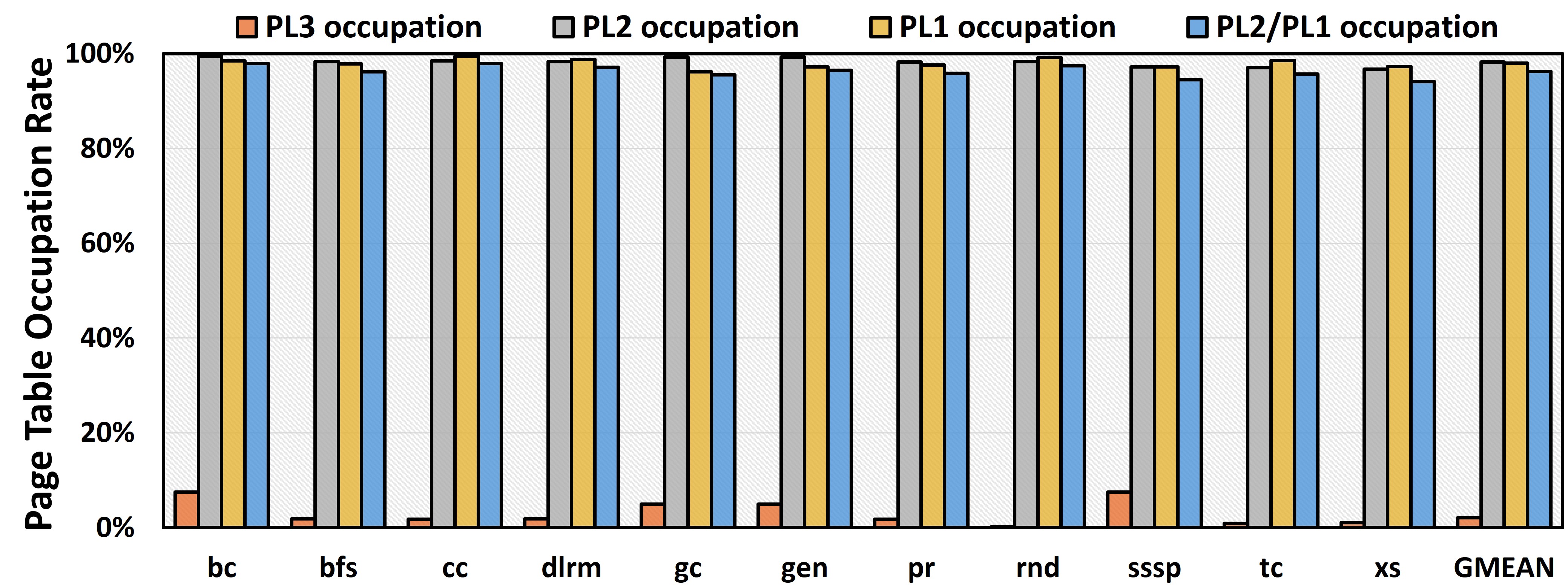}}
    \caption{The page table occupation rates at PL1, PL2, PL3, and combined PL2/PL1 levels for different benchmarks.}
    \label{fig:pt-occu}
\end{figure}
We conduct an analysis of PTE access by examining the cases of address translation on a 4-core NDP system. 
Firstly, we observe that the TLB miss rate reaches 91.27\%, and 65.8\% of the memory accesses are caused by PTEs.
Thus, a significant amount of PTE accesses overwhelms the memory system, starting with the cache hierarchies. 
We evaluate the L1C miss rate of metadata in 4-core NDP systems. 
As Figure~\ref{fig:missrate-cache} shows, the blue bar demonstrates that the average PTE miss rate in the L1 cache reaches 98.28\%. 
We conclude that irregular PTE access cannot effectively utilize the cache in NDP systems. 
What’s worse is that excessive PTE misses will lead to an increase in normal data misses. 
As Figure~\ref{fig:missrate-cache} shows, comparing the green bars and orange bars, we observe that the L1C miss rate of the normal data in the NDP 4-core system reaches 35.89\%, which is 1.37x higher than the 26.16\% in the ideal NDP 4-core system with no address translation.
As a result, the main memory access caused by PTE in the NDP system increases by 200.4x compared to the CPU 4-core system. 
To conclude, our first key observation is that the memory access pattern of PTE metadata in an NDP system is \textbf{extremely irregular}, which results in the following: 
(1) the memory access of metadata cannot benefit from the cache, and 
(2) significantly pollutes the cache, negatively impacting the performance of accessing normal data.

\subsection{Fully Occupied L2 and L1 Page Table}
\label{sec:ob-fullyL21}

The key benefit of the four-level radix page table mechanism in the conventional system is its flexibility, which results in memory space savings, i.e., the next level of the page table is allocated \textbf{only when needed}. 
However, we conduct a study on the occupancy of four-level page tables in the NDP 4-core system. 

As Figure~\ref{fig:pt-occu} shows, the average occupancy rate of the L2 and L1 page tables reaches 98.24\% and 97.97\%, respectively, significantly higher than 0.43\% and 3.12\% of L4 and L3. 
In other words, the full PL2 and PL1 are \textbf{always needed} in NDP systems. 
The flexibility of the 4-level radix page table is overkill in the last 2 levels of fully occupied page tables, and the additional levels increase sequential accesses to the page tables. 
To conclude, our second key observation is that in NDP systems, \textbf{fully occupied} L2 and L1 page tables negate the necessity of maintaining the 4-level radix-tree structure between them.  

These two key observations lead us to two ideas for designing efficient address translation mechanisms for NDP systems.

\section{NDPage Design}
In this section, we describe the mechanisms of NDPage, including an L1 cache-bypass mechanism for metadata and a novel page table architecture with the flattened L2/L1 page table.

\subsection{L1 Cache-Bypass Mechanism for Metadata}

NDPage implements a cache-bypass mechanism only for metadata, i.e., the PTEs, in NDP systems.
This mechanism arises from the observation that highly irregular metadata cannot benefit from the cache in NDP systems, as analyzed in Section~\ref{sec:irregular-pte}.
Therefore, NDPage bypasses the PTEs from the NDP cache, directly accessing the main memory.
The hardware implementation of the cache-bypass mechanism is based on well-established techniques~\cite{mittal2016survey,chi1989improving}. 
We reference these prior works to implement the technology in the hardware design of NDP systems. 
The key focus is on the careful recognition and selection of data to bypass. 
We modify the operating system to mark the regions of the PTEs, which is essentially a 4KB memory, i.e., $2^9$ entries $\times$ 64 bits/entry, in the 4-level radix-tree page table design.
Since 4KB is divisible by 64B, which is the conventional cache line size, making these regions 64B aligned ensures this mechanism does not affect normal data accesses within the cache.
In the NDPage design, for each access to the metadata, the OS triggers a special load instruction to look up the PTEs, depending on the architecture. 
For instance, the x86 ISA provides the PFLD (pipelined floating-point load)~\cite{atkins1991performance} instruction to issue the memory request that bypasses the cache system.
The cache bypassing violates the assumption of inclusion with inclusive multi-level cache hierarchies~\cite{li2012optimal}.
However, since there is only one level of cache in the NDP systems, the violation of the cache bypassing to the inclusive multi-level caches doesn't exist.

\subsection{Flattened L2/L1 Page Table}

\begin{figure}[htbp]
    \centering
    \centerline{\includegraphics[width=\linewidth]{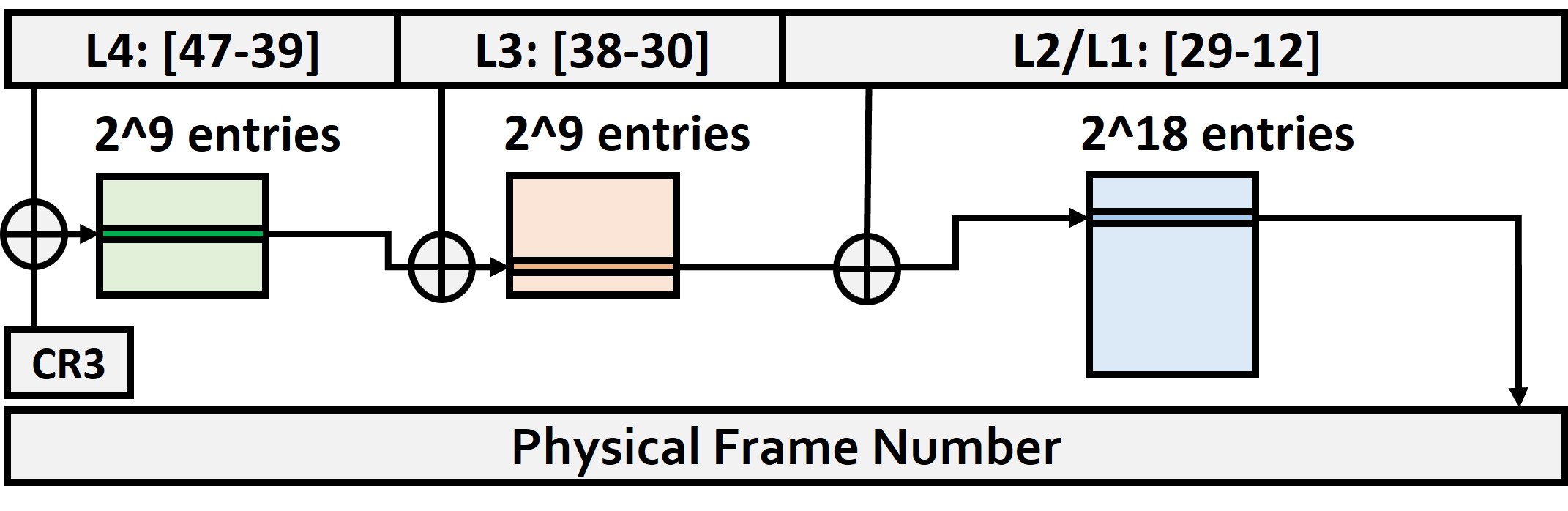}}
    \caption{PTE access flow with new page table design in NDPage.}
    \label{fig:flat-pt}
\end{figure}

The conventional radix page table is organized as a 4-level structure for high flexibility.
However, as shown in Section~\ref{sec:ob-fullyL21}, the 4-level structure is excessive for data-intensive NDP applications since the last two levels of the page table are nearly fully occupied.
Based on this observation, we propose a flattened L2/L1 page table.
Flattening uses the radix nature of the page table to naturally merge levels into single, larger levels, resulting in a shallower tree.
We merge the last two levels of the page table to reduce the number of sequential memory accesses for each PTW from 4 to 3.
Figure~\ref{fig:flat-pt} illustrates the address translation in NDPage with the flattened L2/L1 page table.
By combining the L2 and L1 levels, a single 2 MB flattened node replaces each L2 node and its 512 L1 child nodes, encompassing all $2^9 \times 2^9 = 262,144$ entries.
Although this increases the size of individual page table entries, the overall impact is minimal due to the small fraction of the page table relative to the actual data size.
To support flattened page tables, minor modifications are needed in the system architecture.
Specifically, control registers and page table entries require a single bit to indicate flattened nodes.
The hardware page walker uses these bits to determine the appropriate index bits from the virtual address for each level.
To implement the flattened page tables, we modify the encoding of the virtual address bits for indexing.
Each flattened L2/L1 table structure comprises $2^9 \times 2^9 = 262,144$ entries, fitting into a single 2 MB page.
Therefore, as shown in Figure~\ref{fig:flat-pt}, 18 bits are required for indexing each combined L2/L1 table.

\subsection{Page Walk Caches in NDPage}
\begin{figure}[htbp]
    \centering
    \centerline{\includegraphics[width=1\linewidth]{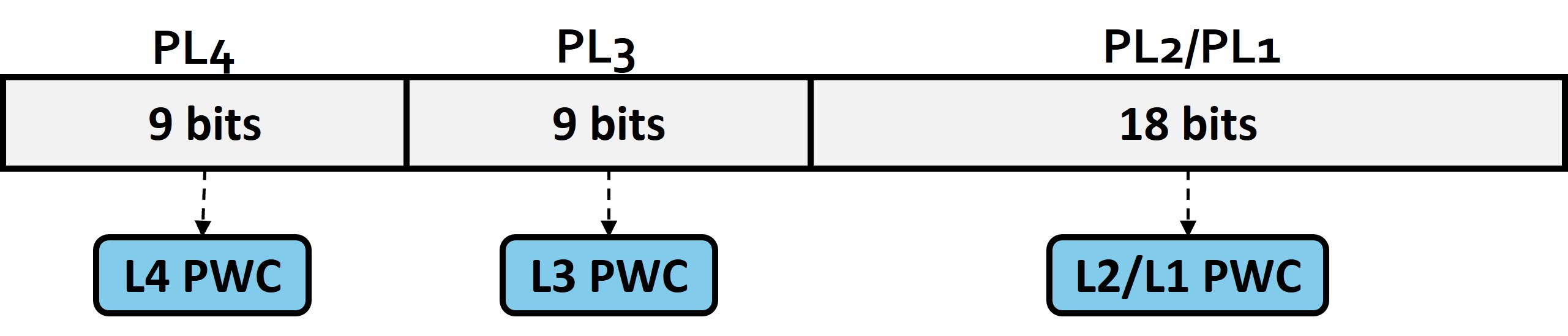}}
    \caption{PWC design in NDPage}
    \label{fig:pwc}
\end{figure}

\begin{figure*}[htbp]
    \centering
    \centerline{\includegraphics[width=0.8\linewidth]{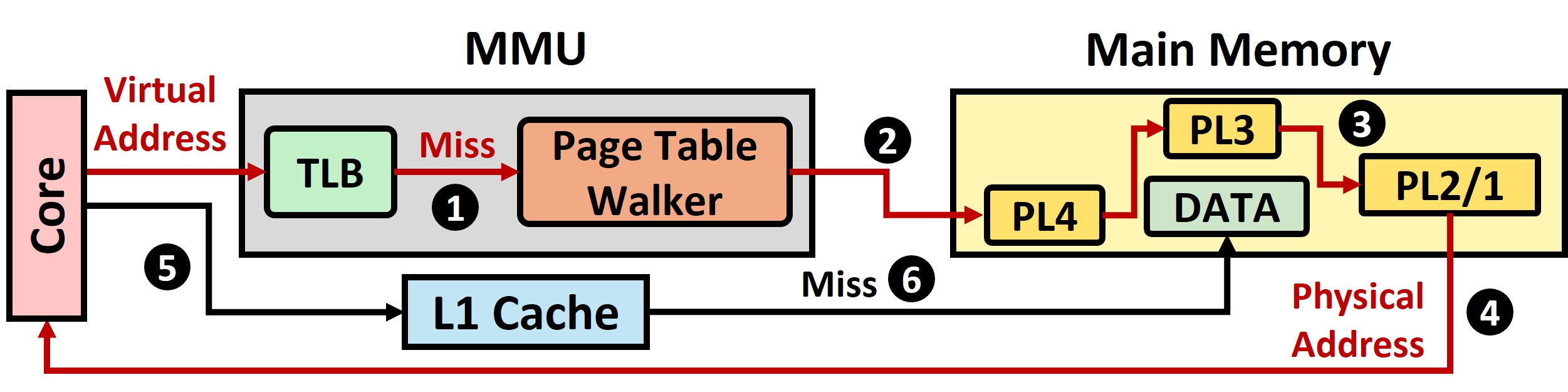}}
    \caption{Address translation and data access flow in an NDP system with NDPage. Red lines indicate the metadata access flow and black lines indicate the normal data access flow.}
    \label{fig:ndpage}
\end{figure*}

The Page Walk Cache (PWC)~\cite{barr2010translation} is a small dedicated cache in modern processors that accelerates address translation by storing recently accessed Page Table Entries (PTEs). 
In a modern processor, each level of the page table has its own PWC. 
We also implement the PWC in the NDPage design to further optimize PTE access. 
We observe that the hit rates of the PWCs in L4 and L3 are nearly 100\% and 98.6\%, respectively, while the hit rates in L2 and L1 are relatively low, averaging 15.4\% in our evaluation. 
This result is consistent with previous works \cite{kanellopoulos2023victima,barr2010translation,kanellopoulos2023utopia}, which also report very high PWC hit rates in L4 and L3 and low hit rates in L2 and L1. 
We conclude that NDPage flattens the last two levels of the page table while retaining the advantages of PWCs in L3 and L4. 
By restricting the high PWC miss rate to a flattened PL2/PL1, instead of two separate PL2 and PL1 levels, NDPage significantly reduces memory accesses.

\subsection{Address Translation Workflow with NDPage}

Figure~\ref{fig:ndpage} demonstrates the address translation and corresponding data access flow in an NDP system with NDPage. 
The core initiates a virtual-to-physical address translation via the MMU, which first incurs a TLB lookup.
When a TLB miss occurs \circled{1}, the MMU triggers a Page Table Walk (PTW) to look up the page tables in the main memory.
Unlike normal data accesses, the metadata accesses caused by the PTW bypass the first-level cache and begin directly at the main memory \circled{2}. 
The PTW requires three sequential page table accesses: first to the PWCs, and if there is a miss, then to the main memory, progressing from the L4 page table to L3, and finally to the flattened L2/L1 page table \circled{3}.
In the end, the page table walker fetches the physical address \circled{4}. 
Once the address translation is complete, the NDP core initiates the normal data access using the physical address at the L1 cache \circled{5}, and if the L1 cache misses, it proceeds to access the main memory \circled{6}.

\section{Evaluation Methodology}
\label{sec:methodology}

We implement NDPage, including the L1 cache-bypass mechanism for metadata, the flattened L2/L1 page table, and the corresponding PWC design using Victima, the simulator used by a prior related work~\cite{kanellopoulos2023victima, kanellopoulos2023utopia}. 
Table~\ref{tab:simconfig} summarizes the system configuration in CPU and NDP systems.

\definecolor{SoftPeach}{rgb}{0.937,0.901,0.901}

\begin{table}[htbp]
\centering
\scriptsize
\caption{Simulation Configuration}
\label{tab:simconfig}
\begin{tblr}{
    width = \linewidth,
    colspec = {Q[95, c]Q[250]Q[250]},
    row{1} = {SoftPeach,c},
    cell{2}{2} = {c=2}{0.6\linewidth, c},
    cell{3}{1} = {r=2}{}, 
    cell{3}{2} = {c=2}{0.6\linewidth, c},
    cell{5}{1} = {r=2}{}, 
    cell{5}{2} = {c=2}{0.6\linewidth, c},
    cell{6}{2} = {c=2}{0.6\linewidth, c},
    cell{7}{2} = {c=2}{0.6\linewidth, c},
    vlines,
    hline{1-20} = {-}{},
  }
  \textbf{system} & \textbf{CPU} & \textbf{NDP}\\
  \textbf{Core} & 1/4/8 x86-64 2.6GHz core(s) & \\
  \textbf{Cache} & L1I/D: 32KB, 8-way, 4-cycle latency \\ &{L2: 512KB, 16-way, 16-cycle latency \\ L3: 2MB/core, 16-way, 35-cycle latency}&{No L2 \\ No L3} \\
  \textbf{MMU} &{L1 ITLB: 128-entry, 4-way, 1-cycle latency \\ L1 DTLB: 64-entry, 4-way, 1-cycle latency} \\ & L2 TLB: 1536-entry, 12-cycle latency \\
  \textbf{Interconnect} & Mesh, 4 cycle hop latency, 512-bit link width \\
  \textbf{Memory} & DDR4-2400, 16GB & HBM2\cite{hbm2}, 16GB \\
\end{tblr}
\end{table}

\begin{table}[htbp]
  \centering
  \scriptsize
  \caption{Evaluated Workloads}
  \begin{tabularx}{\linewidth}{|>{\raggedright\arraybackslash}p{9em}|>{\raggedright\arraybackslash}p{17em}|r|}

    \toprule
    \textbf{Suite} & \textbf{Workload} & \textbf{Dataset size} \\ \midrule
    GraphBIG~\cite{nai2015graphbig} & Betweenness Centrality (BC), Breadth-first search (BFS), Connected components (CC),
     Coloring (GC), PageRank (PR), Triangle counting (TC), Shortest-path (SP) & 8 GB \\ \midrule
    XSBench~\cite{tramm2014xsbench} & Particle Simulation (XS)   & 9 GB \\ \midrule
    GUPS~\cite{plimpton2006simple}  & Random-access (RND)   & 10 GB \\ \midrule
    DLRM~\cite{naumov2019deep}  & Sparse-length sum (DLRM)   & 10 GB \\ \midrule
    GenomicsBench~\cite{subramaniyan2021genomicsbench}  & k-mer counting (GEN)   & 33 GB \\
    \bottomrule
    \end{tabularx}
    \label{tab:workloads}
\end{table}

\textbf{Workloads.} Table~\ref{tab:workloads} shows all benchmarks we use to evaluate NDPage. 
We choose 11 data-intensive applications compatible with NDP architecture which are also used in previous works about address translation~\cite{skarlatos2020elastic,park2022every,ainsworth2021compendia,gupta2021rebooting}. 
We simulate the execution of 500M instructions per core.

\textbf{Evaluated Address translation Mechanisms.} 
We evaluate six different mechanisms in NDP systems:
(1) Radix: Conventional x86-64 system with a four-level radix-based page table.
(2) Elastic Cuckoo Hash Table (ECH)~\cite{skarlatos2020elastic}: the state-of-the-art hash-based page table.
ECH increases the parallelism of PTE access and reduces PTW latency. 
(3) Huge Page~\cite{corbet2011}: the operating system allocates memory in 2MB chunks.
(4) Ideal: every address translation request hits the L1 TLB, and the access latency to the L1 TLB is zero. 
This scenario denotes the performance limit of address translation mechanisms.
(5) NDPage: this work.

\section{Evaluation Results}
\label{sec:results}

\subsection{Performance in the Single-Core System}
\begin{figure}[htbp]
    \centering
    \centerline{\includegraphics[width=\linewidth]{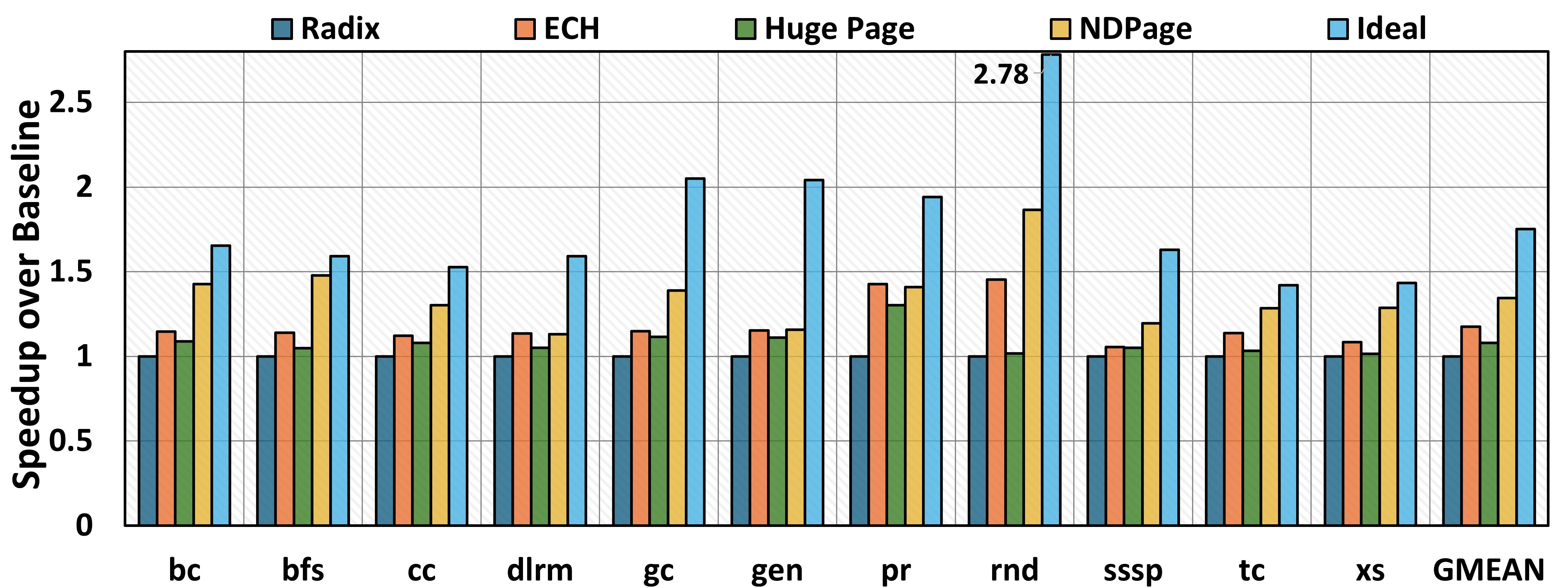}}
    \caption{Speedup provided by ECH, Huge Page, NDPage, and Ideal over Radix in single-core execution.}
    \label{fig:ndpage-1core}
\end{figure}
Figure~\ref{fig:ndpage-1core} shows the speedup provided by the evaluated address translation mechanisms compared to the Radix baseline in single-core NDP systems.
We make two key observations based on the results:
(1) On average, NDPage outperforms the second-best address translation mechanism (ECH) by 14.3\% and Radix by 34.4\%, demonstrating NDPage's capability to mitigate the address translation overhead in the NDP architecture.
(2) Although Huge Page also requires only three-level page tables, reducing one sequential page table access compared with Radix, NDPage outperforms Huge Page by 24.4\%, highlighting the performance advantage of NDPage's Metadata Bypass mechanism.

\subsection{Performance in the Multi-Core System}
\begin{figure}[htbp]
    \centering
    \centerline{\includegraphics[width=\linewidth]{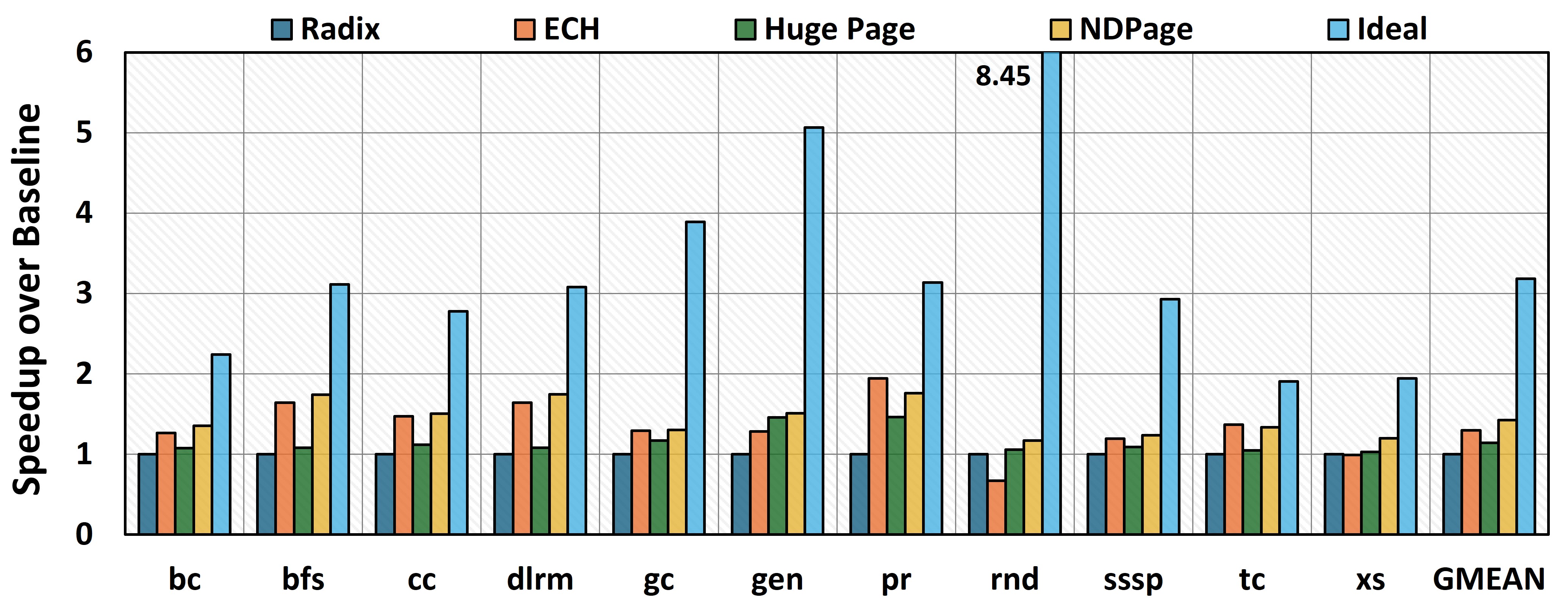}}
    \caption{Speedup provided by ECH, Huge Page, NDPage, and Ideal over Radix in 4-core execution.}
    \label{fig:ndpage-4core}
\end{figure}
\begin{figure}[htbp]
    \centering
    \centerline{\includegraphics[width=\linewidth]{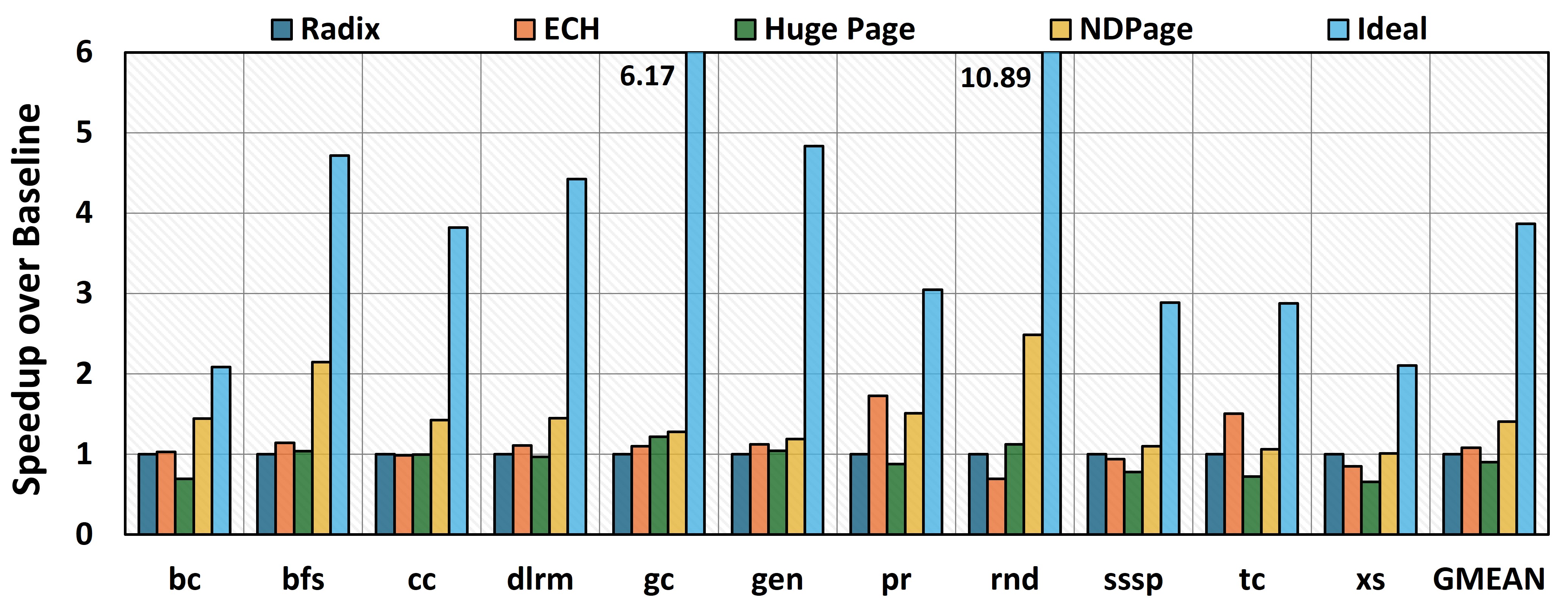}}
    \caption{Speedup provided by ECH, Huge Page, NDPage, and Ideal over Radix in 8-core execution.}
    \label{fig:ndpage-8core}
\end{figure}

Figure~\ref{fig:ndpage-4core} and Figure~\ref{fig:ndpage-8core} show the execution speedup provided by the evaluated address translation mechanisms in 4-core and 8-core NDP systems, respectively. 
We make several key observations based on the results: 
(1) on average, NDPage outperforms the second-best address translation mechanism (ECH) by 9.8\% and 30.5\% in 4-core and 8-core NDP systems, respectively, demonstrating NDPage's significant performance benefits and scalability with NDP multi-core architectures.
(2) In the 8-core system, Huge Page achieves only 90.1\% of the performance of the Radix baseline. 
We conclude that as the workload scale and the number of NDP cores increase, Huge Page brings increased page fault latency, bloat memory footprint, and rapid consumption of available physical memory contiguity\cite{kwon2016coordinated}. 
As a result, Huge Page leads to performance degradation. 
However, based on our second key observation, NDPage reduces sequential page table accesses while maintaining the flexibility of a 4KB page, performing well (56.2\% speedup compared to Huge page on average) in the 8-core NDP system.

\section{Related Work}
To our knowledge, NDPage is the first work to provide a highly efficient, programmer-transparent virtual-to-physical translation mechanism for NDP systems, achieved through a novel page table design with modest hardware modifications. 
Several prior works have focused on reducing address translation overhead in NDP systems. 
vPIM~\cite{fatima2023vpim} optimizes page table accesses using a \textit{network-contention-aware hash} page table and allocates some NDP cores for \textit{pre-translation}. 
However, the hash-based page table limits certain virtual memory functionalities, such as page data sharing, and requires programmers to modify their applications to conform to the vPIM framework. 
DIPTA~\cite{picorel2017near} improves page table access by strategically placing page tables closer to the data. 
Nevertheless, DIPTA restricts page mapping associativity, potentially leading to significant performance degradation in applications that frequently encounter page conflicts. 
In contrast, NDPage provides a programmer-transparent solution that imposes no restrictions on page mapping.

\section{Conclusion}
NDP architecture makes efficient use of high memory bandwidth and is compatible with data-intensive workloads. 
However, data-intensive workloads experience irregular memory accesses and long-latency page table walks, leading to excessive overheads in address translation using the conventional table mechanism in NDP systems. 
We propose NDPage, an address translation mechanism that (1) bypasses the L1 cache in NDP systems when accessing PTEs, and (2) flattens the fully occupied L2 and L1 page tables. 
Our evaluation demonstrates that NDPage provides substantial performance improvements in NDP systems, offering an efficient address translation mechanism.

\section*{Acknowledgment}
This work is partly supported by Strategic Priority Research Program of the Chinese Academy of Sciences (Grant No. XDB0500102). The computing resources are financially supported by Laoshan Laboratory (LSKJ202300305).





\bibliographystyle{abbrv}
\bibliography{conference_101719}

\end{document}